# Development and Comprehensive Evaluation of TMR Sensor-Based Magnetrodes

*Jiahui Luo [1,2], Zhaojie Xu[1,2], Zhenhu Jin [1,2], Mixia Wang[1,2],\*, Xinxia Cai[1,2,]\* and Jiamin Chen [1,2,3,]\**

[1]State Key Laboratory of Transducer Technology, Aerospace Information Research Institute, Chinese Academy of Sciences, Beijing 100190, China

[2]School of Electronic, Electrical and Communication Engineering, University of Chinese Academy of Sciences, Beijing 100049, China

[3]College of Materials Sciences and Opto-Electronic Technology, University of Chinese Academy of Sciences, Beijing 100049, China

\*    Correspondence: chenjm@aircas.ac.cn, wangmixia@mail.ie.ac.cn, xxcai@mail.ie.ac.cn

**Abstract**：Due to their compact size and exceptional sensitivity at room temperature, magnetoresistance (MR) sensors have garnered considerable interest in numerous fields, particularly in the detection of weak magnetic signals in biological systems. The "magnetrodes", integrating MR sensors with needle-shaped Si-based substrates, are designed to be inserted into the brain for local magnetic field detection. Although recent research has predominantly focused on giant magnetoresistance (GMR) sensors, tunnel magnetoresistance (TMR) sensors exhibit significantly higher sensitivity. In this study, we introduce TMR-based magnetrodes featuring TMR sensors at both the tip and mid-section of the probe, enabling detection of local magnetic fields at varied spatial positions. To enhance detectivity, we have designed and fabricated magnetrodes with varied aspect ratios of the free layer, incorporating diverse junction shapes, quantities, and serial arrangements. Utilizing a custom-built magnetotransport and noise measurement system for characterization, our TMR-based magnetrode demonstrates a limit of detection (LOD) of 300pT/$\sqrt{Hz}$ at 1 kHz. This implies that neuronal spikes can be distinguished with minimal averaging, thereby facilitating the elucidation of their magnetic properties.

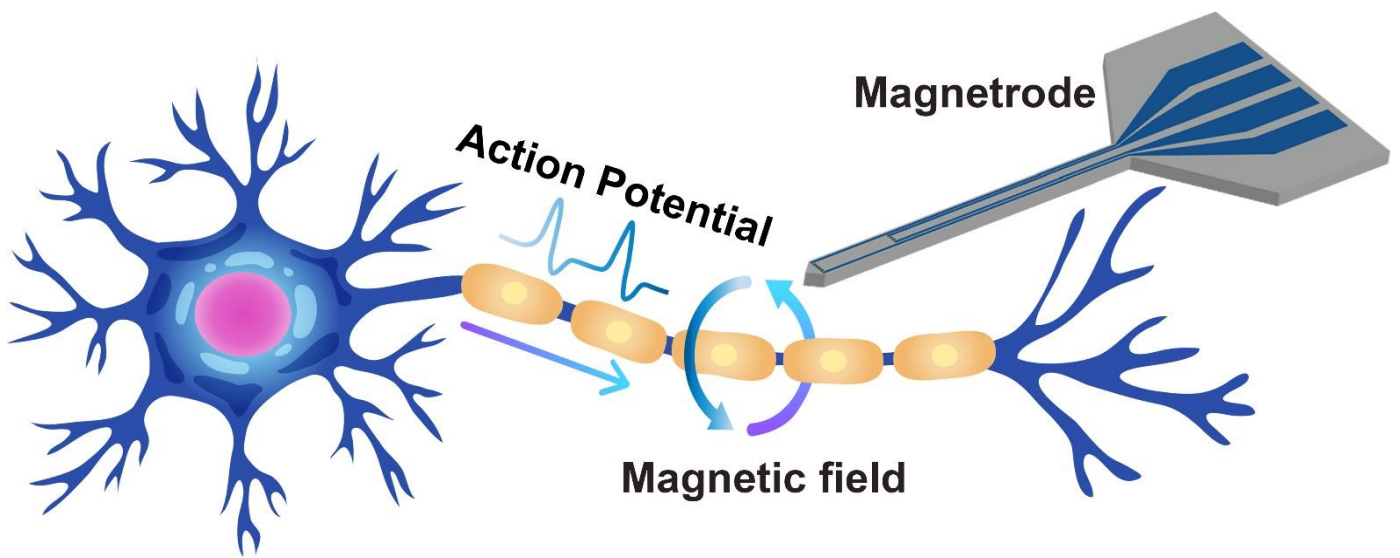

## INTRODUCTION

The neural network of the human brain comprising hundreds of millions of neurons[1], remains a subject with a complex working process that is yet to be fully understood. During information transmission, neurons not only generate the well-known electric current but also produce a weak magnetic field. In comparison to electrical signals, magnetic fields offer several advantages: they are non-contact, undergo no distortion during propagation in the brain, are reference-free, and vector signals. Full-scale detection of this magnetic field, which holds potential to enhance our understanding of the brain, is garnering increasing interest.

At the macroscopic level, magnetoencephalography (MEG), utilizing superconducting quantum interference devices (SQUIDs), enables ultra-high sensitivity brain imaging[2]—a crucial tool in identifying lesions associated with epilepsy and other disorders. Microscopically, the nitrogen-vacancy center technique[3] has been successful in detecting magnetic fields generated by individual neurons. However, the limitations of SQUID and nitrogen-vacancy centers, including their bulky size, high fabrication costs, and complex mechanisms, restrict their utility in in vivo local magnetic field recording.

Spintronics-based magnetoresistance (MR) sensors are regarded as promising candidates for in vivo local magnetic recording, owing to their high sensitivity, compact size, and low power consumption. Utilizing microelectromechanical systems (MEMS) technology, micron-sized MR sensors have been integrated onto the tips of sharp needle-shaped Si-based probes, facilitating easy insertion for local magnetic field detection at various positions. These innovative neural probes are termed "magnetrodes"[4], a nomenclature inspired by electrodes. Recently, magnetrodes employing giant magnetoresistance (GMR) sensors have been utilized for magnetic recording both in vitro[5-7] and in vivo[4, 8]. More detailed information can be found in the review[9] previously published by the author.

We anticipate that magnetrodes will not only detect local magnetic fields but also differentiate the direction and magnitude of fields generated by single neuron spikes, significantly advancing research in biological magnetism and neuroscience. According to Biot-Savart law, the magnetic field amplitude from a single neuron spike does not exceed 100pT at approximately 1kHz frequency, necessitating enhanced sensor performance. The tunnel magnetoresistance (TMR) sensor, a key MR sensor variant, operates based on spin-electron tunneling through an ultra-thin insulating layer. TMR sensors, in comparison to GMR sensors, demonstrate markedly higher sensitivity and have been effectively utilized in biomagnetic field detection, including magnetocardiography (MCG)[10, 11] and magnetoencephalography (MEG)[12]. Nevertheless, research on TMR-based magnetrodes[13] remains limited, partly due to the intricate fabrication process of TMR sensors and their relatively higher 1/f noise at low frequencies.

In this study, we present a magnetrode design featuring a probe width of 300μm and an angle of 80°, with TMR sensors strategically positioned at the tip and mid-section of the probe for detecting local magnetic fields at varied locations. Crucially, a series-connected architecture was implemented to minimize the TMR sensor's output noise. Additionally, we optimized the free layer's aspect ratio, junction shapes, and array configurations to enhance MR ratio, reduce hysteresis, and improve sensitivity along with other magnetotransport properties. Moreover, we developed a bespoke magnetotransport and noise measurement system, conducting comprehensive evaluations of its performance characteristics. The characterization of our fabricated TMR-based magnetrodes revealed a sensitivity of 22.29%/mT and a LOD of $300pT/\sqrt{Hz}$ at 1kHz, surpassing the performance of the best-known GMR-based magnetrodes[7] by nearly an order of magnitude. Consequently, employing our advanced TMR-based magnetrodes enables faster resolution of magnetic signals from single neuron spikes, requiring fewer averages. Finally, the magnetic signals detected by the magnetrodes in the in vitro simulation experiments exhibited firing rate of action potential spikes consistent with the electrical signals, preliminarily demonstrating the feasibility of detecting action potentials.

## RESULTS AND DISCUSSION

**TMR multilayer structures.** The TMR stack in this study is characterized by a dual-pinning structure, composed of $Si/SiO_2$/Ta/Ru/Ta/Ru/Ta/Ru/PtMn/CoFe/Ru/CoFeB/MgO/CoFeB/NiFe/IrMn/NiFe/IrMn/Ru/Ta/Ru. As illustrated in Figure 1a, the free and reference layers are individually pinned by IrMn and PtMn antiferromagnetic layers, respectively. The implementation of synthetic antiferromagnetic (SAF) structures aims to stabilize the magnetization direction of the reference layer, concurrently minimizing the stray magnetic field's impact on the free layer[14]. The composite free layer, consisting of CoFeB/NiFe/Ru/IrMn, effectively reduces the saturation field while simultaneously achieving a higher TMR ratio, leading to enhanced sensitivity[15]. Thanks to the different Néel temperature of the two antiferromagnetic layers, it is possible to orthogonalize the easy axes of the free and pinned layers by means of two annealing steps before and after microfabrication, which leads a linear response. The wafer undergoes dual annealing processes by Vacuum Magnetic Annealing Furnace (Model F800-35/EM7, East Changing Technologies, China) to achieve a high MR ratio and linear response.

**Development and design of Magnetrodes.** In biological applications, the optimization of magnetic tunnel junction (MTJ)-based TMR sensors for weak biological magnetic fields, typically below several kHz, is crucial. Achieving lower noise levels is imperative for improved detectivity alongside the continuous enhancement of the MR ratio. Previous studies have shown that utilizing an MTJ structure in an array configuration effectively suppresses 1/f noise[16, 17], enhancing LOD. Our design, inspired by Jin et al.[18], features a similar series structure where MTJ junctions share a common bottom electrode and are connected serially through deposited gold electrodes. We explored hysteresis-free linearized output by developing magnetrodes with various MTJ junction shapes, including rounded rectangle and ellipse. Investigations by MARM, utilizing micromagnetic simulations for device bit shape optimization, indicated that elliptical junctions effectively reduce demagnetization energy at the edges[19-21]. To maintain a linear response, we regulated the aspect ratio of the device to ensure perpendicularity between the easy axes of the reference and free layers[22], as per the Stoner-Wohlfarth model[23]. Additionally, a dual-pinning structure[15], involving an antiferromagnetic pinning layer grown on the free layer, facilitated perpendicular pinning directions. As depicted in Figure 1b, different configurations of magnetrodes, classified as parallel or orthogonal based on sensor alignment, underwent comparative performance analysis considering free-layer aspect ratio, MTJ junction shapes, and serial connections. The placement of sensor arrays at both the tip and middle of the probe aimed to enhance the detection of localized magnetic fields at various positions. Figure 1c illustrates a schematic diagram of magnetrodes detecting a magnetic field, and Figure 1d gives a scanning electrode microscope (SEM) of an orthogonal magnetrode with rounded rectangular junctions.

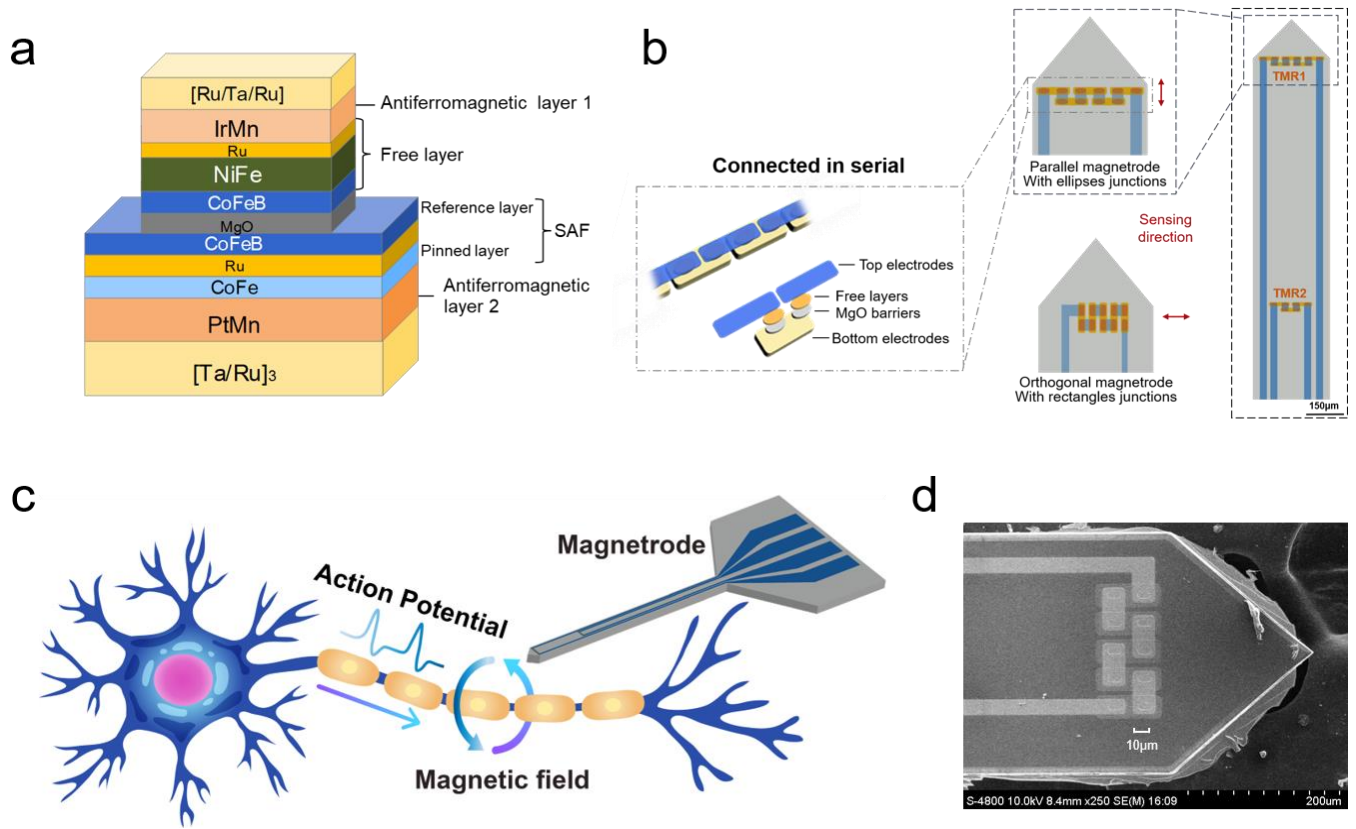

**Figure 1.** (a) Magnetic film structure; (b) Schematic view of the magnetrodes, highlighting the structure of connected in serial and the placement of TMR1 and TMR2 sensors and detailing the geometric variances in sensing direction relative to the probe direction in two distinct configurations: one featuring a sensing direction parallel to the probe with elliptical MTJ junctions, and another with an orthogonal sensing direction accompanied by rounded rectangular MTJ junctions; (c) Schematic diagram of magnetrode to detect magnetic fields generated by action potentials; (d) SEM image of the tip of an orthogonal magnetrode with rounded rectangular junctions.

**Magnetotransport Performance of Magnetrodes.** To meet the need of detecting weak magnetic fields, we explore the best-performing magnetrodes with high sensitivity and low hysteresis as our goal for output characteristics. Initially, the magnetotransport performance of magnetrodes, with a short axis length of 15μm and aspect ratios of 2, 3, 4, was compared. Figure 2a presents the R-H output curves of MTJs with different free layer sizes and different junction shapes. The insets on the left and right depict the variation of hysteresis and sensitivity with the aspect ratio of the free layer for different junction shapes, respectively. Different symbols represent different junction shapes. It is worth noting that although the junction shapes are different, they exhibit the same variation pattern.

First focusing on the R-H output curves of MTJs with different free layer sizes, it can be observed that as the aspect ratio and area of the free layer increase, the TMR ratio decreases. RA value refers to the product of the resistance and area of the TMR multilayer film, and remains constant of the same film layer. Therefore, as the aspect ratio of the free layer increases, the junction area also increases, leading to a decrease in the effective resistance of the device. On the other hand, the parasitic resistance introduced by the gold electrode contact remains almost constant during this process. Consequently, the proportion of parasitic resistance in the total resistance gradually increases, resulting in a decrease in the TMR value. The decrease in TMR further induces changes in sensitivity, as shown in the inset on the right side of Figure 2a. Sensitivity is determined by $TMR/2H_k$, where $H_k$ is a magnetic anisotropy of the FL[18]. An increase in aspect ratio signifies an increase in the anisotropy field, which also leads to a decrease in sensitivity.

Hysteresis is another important indicator for sensor applications, as shown in the inset on the left side of Figure 2a, which is negatively correlated with the aspect ratio of the free layer. In our designed structure, the pinning layer's easy magnetization axis remains stable along the device's short-axis direction, owing to its coupling with the antiferromagnetic and SAF layers. The free layer's easy magnetization axis progressively aligns with the device's long-axis direction, influenced by shape anisotropy. MTJs with a high FL aspect ratio tend to exhibit a high anisotropy field due to geometric effects. This causes the FL magnetic moments to be preferentially aligned along the long axis of the junction, making it easier to be orthogonal to the pinning direction, thereby reducing the device's hysteresis.

Achieving both high sensitivity and low hysteresis in sensors is not an easy task. As evident from the two insets in Figure 2a, there is a certain degree of trade-off between sensitivity and hysteresis. With an increase in the aspect ratio of the free layer, the output hysteresis decreases, but at the same time, sensitivity also decreases. Therefore, we need to strike a balance between the two, and the aspect ratio of 3 for the junction area is considered to be an appropriate choice. This choice allows us to maintain relatively high sensitivity while obtaining lower hysteresis, without excessively affecting spatial resolution due to excessive footprint.

Subsequently, attention is directed towards the junction shape of MTJ. Figure 2b displays the R-H curves for both rounded rectangular and elliptical junctions, featuring a free layer aspect ratio of 3 and a series connection of 12. In line with expectations, the output hysteresis of elliptical MTJ junctions is significantly reduced with sensitivity and linearity remaining largely unchanged. The elliptical shape distributes the magnetic domain more uniformly and reduces the sharp corners where demagnetization energy tends to be concentrated. This leads to a more stable magnetic state and a more linear response from the sensor. Furthermore, the elevated MR value of elliptical junctions can be attributed to their larger effective area for magnetic domain flipping.

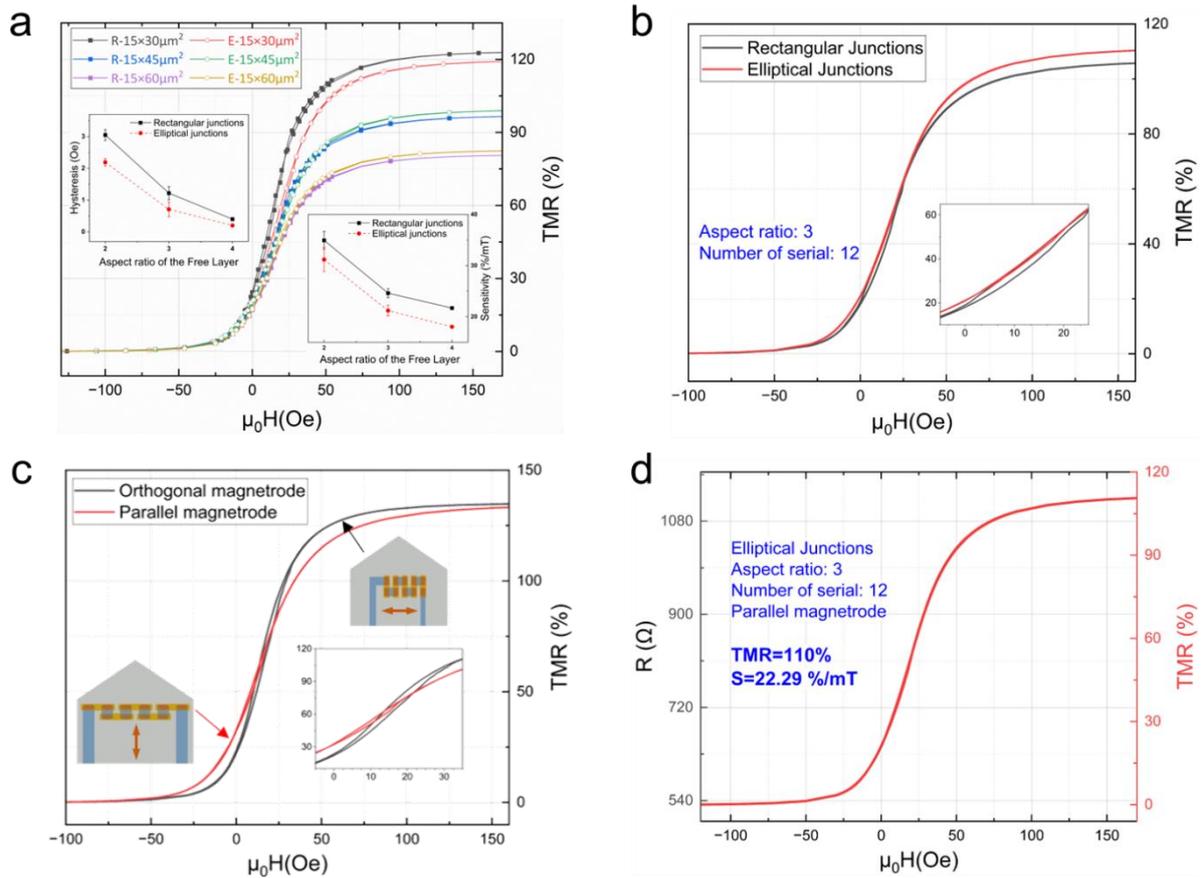

**Figure 2.** (a) R-H output curves of MTJs with different free layer sizes and different junction shapes. Two insets on the left and right show the variations of hysteresis and sensitivity with the aspect ratio of the free layer for different junction shapes. R-rounded rectangular junction, E-elliptical junction, different symbols represent different junction shapes in the insets; (b) R-H curves of magnetrodes with rounded rectangular and elliptical junctions are depicted, with the aspect ratio of the free layer being 3 and the number of serial connections being 12 for both configurations; (c) R-H curves for different array configurations of elliptical junctions, with an aspect ratio of 2 and a series connection of 10. A parallel magnetrode refers to the sensor's sensing direction being parallel to the probe direction, while an orthogonal magnetrode indicates that the two directions are orthogonal; (d) R-H curves of the optimally performing magnetrode featuring elliptical junctions. This magnetrode is characterized by a free layer with an aspect ratio of 3 and a total of 12 serial connections.

Additionally, the impact of array configurations with different sensing directions on magnetrode output was analyzed. Figure 2c presents the R-H curves for various array configurations of elliptical junctions, each having an aspect ratio of 2 and a series connection of 10. The figure indicates that while the MR ratios of both array configurations are nearly identical, the parallel magnetrode demonstrates reduced hysteresis, albeit with a corresponding decrease in sensitivity. One plausible conjecture for the decrease in hysteresis is that in parallel magnetrodes, the long axis of the MTJs aligns with the arrangement direction. In this case, the flipping of the free layer tends to be more uniform and consistent, especially when arranged in a single column (although complete realization is constrained by the width of magnetrodes). On the other hand, in orthogonal magnetrodes, where the long axis of the MTJ is perpendicular to the arrangement direction, the flipping of the free layer in individual junctions may mutually influence each other, potentially leading to larger magnetic hysteresis.

Ultimately, by integrating the earlier analyses of junction shape, free layer aspect ratio, and array configuration, the optimal magnetrode configuration for magnetotransport performance was determined. The optimal magnetrode features an elliptical junction with an aspect ratio of 3, a series connection of 12, and a parallel configuration. Magnetotransport performance results, as depicted in Figure 2d, show that this magnetrode achieves a high TMR ratio of 110% and a sensitivity of 22.29%/mT.

**Noise Measurement System Performances.** Prior to conducting noise measurements on the magnetrodes, it is imperative to evaluate the system's accuracy. Initially, the amplifier's input voltage noise level is determined by shorting its input. The nominal value is $4nV/\sqrt{Hz}$ (as shown by the red line in Figure 3a), and our test results, represented by the black curve in Figure 3a, are in close agreement with this theoretical value. Subsequently, metal film standard resistors with varying resistance values are utilized to assess potential systematic errors, attributed to the test environment and amplifier noise, by measuring their noise levels and comparing them to theoretical values of resistor white noise. The amplifier is set to AC-coupled mode with a gain setting of 5000. By averaging data around the 1 kHz frequency on the noise curve and subsequently subtracting the amplifier's voltage noise influence, the resistor noise value is obtained.

Figure 3b displays noise curves for resistors of 2kΩ, 20kΩ, and 200kΩ. These curves predominantly exhibit horizontal, linear profiles, indicative of the frequency characteristics of white noise. By averaging the data near the 1 kHz frequency on the noise curve, the measured resistor noise values are compared with their theoretical counterparts, as detailed in Table 1. The system error falls within ±5.6%, attesting to the high accuracy of the noise measurement system.

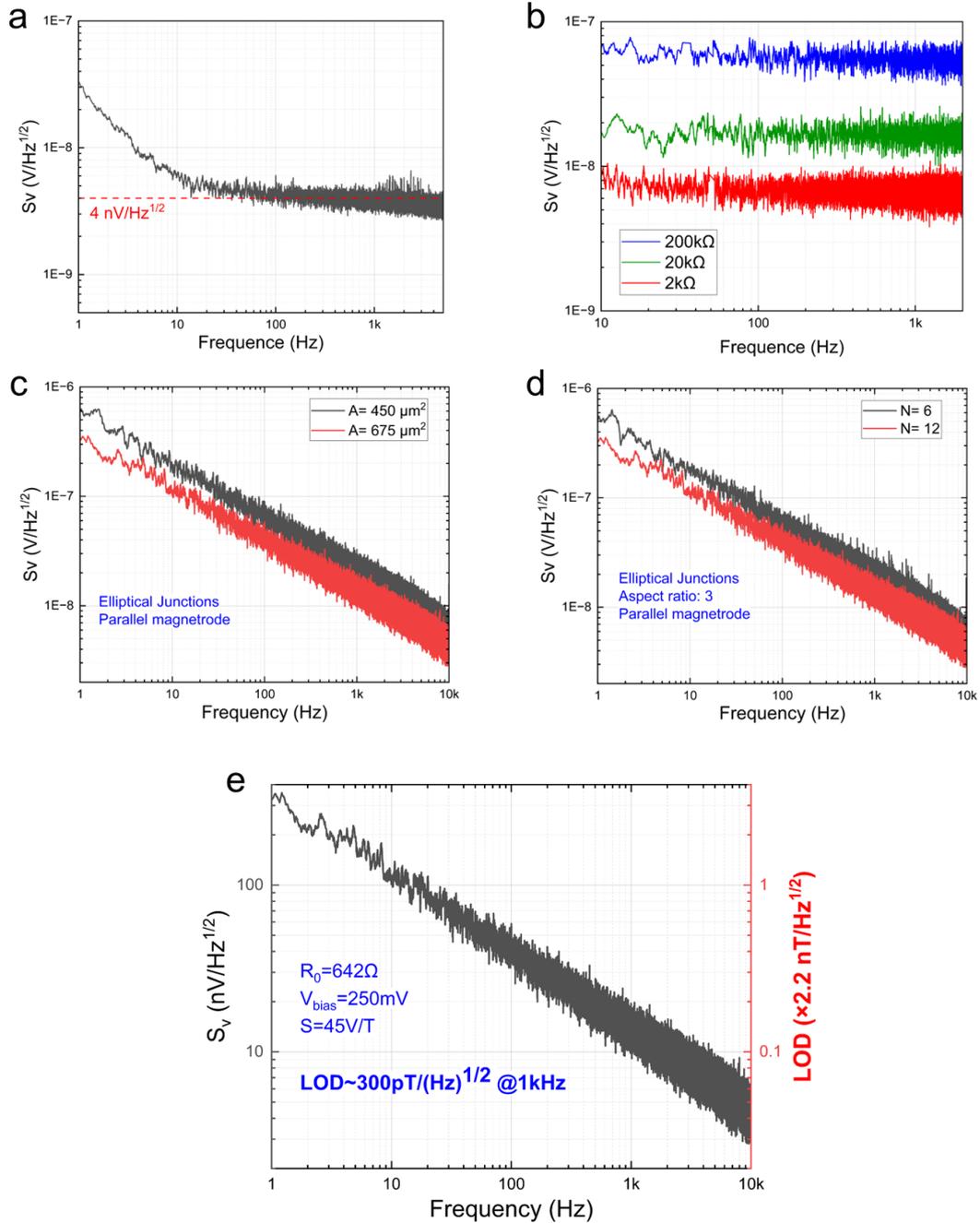

**Figure 3.** (a) Test results of amplifier voltage noise (in black), with the theoretical values represented by the red line; (b) Noise curves of different standard resistors; (c) Noise power spectral density of devices with different junction areas; (d) Noise power spectral density of devices with different number in series; (e) Noise spectral density $S_v$ (left) and LOD (right) as a function of frequency for magnetrode with elliptical junctions. This magnetrode is characterized by a free layer with an aspect ratio of 3 and a total of 12 serial connections.

**Table 1.** Systematic error under standard resistance measurement.

| R (kΩ) | 2 | 20 | 200 |
|---|---|---|---|
| Theoretical value of white noise (nV/$\sqrt{Hz}$) | 5.66 | 17.8 | 56.58 |
| Experimental value of white noise (nV$\sqrt{Hz}$) | 5.76 | 16.8 | 55.63 |
| System error (%) | 1.8 | -5.6 | -1.7 |

**Noise Performance of Magnetrodes.** In magnetic sensors, the Limit of Detection (LOD) is defined as

$$LOD = \frac{S_V}{\Delta V / \Delta H} \quad (1)$$

The LOD is influenced not only by device sensitivity but also by its noise characteristics. Noise in MR devices, based on their physical origins, can be categorized into white noise and low-frequency noise. White noise, frequency-independent in nature, encompasses both thermal noise and shot noise. Low-frequency noise comprises 1/f noise and random telegraph noise (RTN), with the suppression of 1/f noise being crucial in MR device development. Specifically, 1/f noise originates from both electrical and magnetic 1/f noise components. Magnetic 1/f noise is associated with domain hopping between metastable states[24], while electrical 1/f noise arises from defects in the oxide barrier layer or oxide/ferromagnetic interface, and charge trapping effects due to oxygen vacancies[25]. Electrical 1/f noise in MTJs is typically represented by a Hooge-model-like formula:

$$S_V^{elec,1/f} = \frac{\alpha V^2}{Af} \quad (2)$$

where α is the electrical 1/f noise Hooge parameter, V is the bias voltage and A is the junction area. This formula indicates that $S_V^{elec,1/f}$ decreases as the junction area A increases, given a fixed bias voltage. In practice, the noise performance of the magnetrode was assessed using the noise measurement system described before, under conditions devoid of external magnetic fields and with a fixed bias voltage. The experimental results, as demonstrated in Figure 3c, align well with theoretical predictions, showing that magnetrodes with larger junction areas exhibit lower $S_V$ $(V/Hz^{1/2})$.

Additionally, serially connecting MTJs has proven effective in suppressing output noise. This approach specifically reduces the 1/f noise contribution by a factor of $1/\sqrt{N}$ [26], where N is the number of junctions in series. Figure 3d's results confirm that a magnetrode with 12 MTJs in series exhibits lower noise than one with 6 MTJs in series.

Finally, following the comparative analysis of magnetotransport and noise performance, the left and right axes of Figure 3e respectively present the $S_V$ and LOD results for the magnetrode with optimal performance. Averaging values near the 1 kHz frequency yields an LOD of 300pT/$\sqrt{Hz}$ for this magnetrode and a rms noise level within the [800 to 1200 Hz] bandwidth of 6.3nT, which averaged on 4,000 events would lower it down to 100pT. This signifies an improvement nearly an order of magnitude greater than the recently published LOD for a GMR-based magnetrode[7], suggesting the potential for faster detection of magnetic signals from a single neuron with fewer averages.

**In Vitro Simulation Experiments.** The frequency of action potentials, often referred to as neuronal firing rate, characterizes the response of neurons. Due to its relative simplicity and attractiveness, it has been widely used in data analysis and modeling[27]. The firing rate of the magnetic signal obtained experimentally matches the electrical signal, thereby providing preliminary validation of the practicality of TMR-based magnetrodes. Specifically, we utilized a commercial neural signal simulator to generate simulated neuronal signals, where action potentials are emitted in a continuous discharge lasting 1 second every 9 seconds. Within this continuous discharge of 1 second, individual action potentials occur at intervals of 0.03 seconds, resulting in an approximate action potential firing rate of 30 Hz. The electrical signal is applied to the ends of a copper wire with a diameter close to that of the neuron axon, and the current flowing through the copper wire further generates a magnetic field for detection by the magnetrodes.

In the experiment, electrical and magnetic signals were synchronized for recording to directly correspond action potentials in the electrical signals with corresponding events in the magnetic signals. From the collected electrical signals, a series of time points were identified where spikes occurred during periods of intense discharge. Utilizing this data, the collected magnetic signals were subjected to truncation, filtering, FFT transformation, and averaging, ultimately revealing distinct peaks in the frequency domain at around 30Hz, as shown in figure 4. This alignment with the firing rate of action potentials confirms our magnetrodes' capability to detect the magnetic field generated by action potentials.

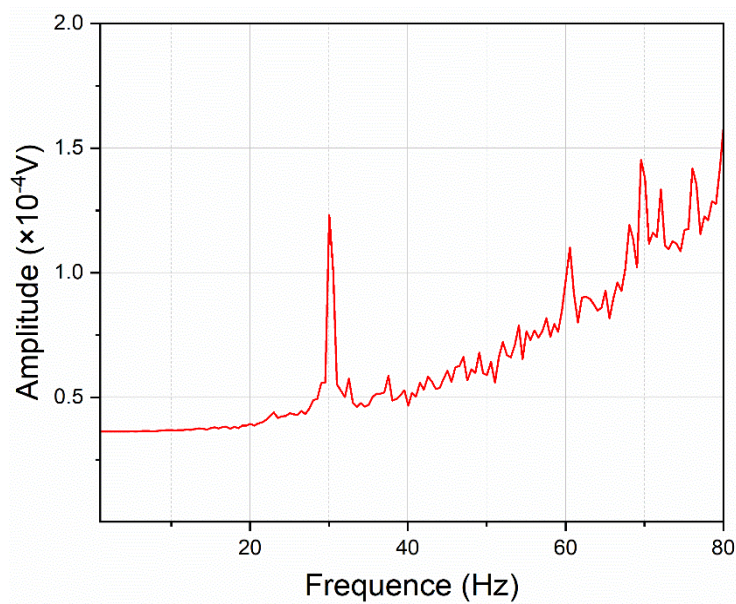

**Figure 4.** Frequency domain plot of the magnetic field corresponding to action potentials detected by magnetrodes.

**CONCLUSIONS**

This study provides a detailed overview of the magnetrodes' manufacturing process and evaluates their overall performance using a specialized magnetotransport and noise measurement system. Prior to these evaluations, calibrating the baseline noise of the measurement systems, especially the noise measurement system, was crucial. Actual test results for voltage noise from the SR560 and thermal noise from a series of constant metal film standard resistors closely matched theoretical values, confirming the effectiveness of our noise measurement system in characterizing device noise.

Following this, we commenced with the performance characterization of the magnetrodes. We began by comparing magnetotransport performances to understand how MTJ junction shape, free layer aspect ratio, and serial array configuration affect the magnetrodes' hysteresis and sensitivity. The results indicate that elliptical junctions can achieve relatively smaller hysteresis. Additionally, the selection of aspect ratio for the free layer needs to strike a balance between sensitivity and hysteresis. The arrangement of junctions in the array can also affect the final output, although its theoretical explanation requires further exploration. We then investigated how increasing the junction area and the number of series connections impact the suppression of low-frequency noise. Remarkably, we achieved an impressive LOD of $300pT/\sqrt{Hz}$ @1kHz, which notably exceeds the metrics previously reported for GMR-based magnetrodes.

To validate the practicality of the magnetrodes, we also conducted simple in vitro simulation experiments. The magnetic signals detected by the magnetrode exhibited firing rate of action potential spikes consistent with the electrical signals, thus preliminarily demonstrating its feasibility for action potential detection. However, conducting in vivo experiments at this stage presents several challenges. Environmental interference, including electromagnetic noise pollution, may be severe enough to mask the desired signal. Additionally, simultaneous recording of electrodes and magnetrodes may introduce additional noise due to mutual interference. In the future, our efforts will focus on integrating electrodes and magnetrodes onto a single probe for in vivo experiments under cleaner magnetic field conditions, such as a magnetic shielding room.

## METHODS

**Fabrication of Magnetrodes.** MTJs are patterned using self-aligned photolithography and a two-step ion beam etching (IBE) process. In the IBE process, the first step defines the pinned layer electrode, and the second step of free junction etching stops in the middle of the MgO barrier, which is crucial for electron tunneling. To enable real-time etching endpoint detection, we upgraded the LKJ-1A-150 ion beam etcher by integrating an optical detector into the expansion chamber for in-situ emission spectrum analysis[28]. This optical detector is linked to an external spectrometer through optical fibers and vacuum connectors. Emission spectra collected are transmitted to a computer, where analysis of characteristic peak intensities aids in determining the etching process endpoint. Figure 5a-b depict the three-dimensional model of the ion beam etcher and the monitored spectra results during the initial etching step. Following the pillar etching, 200 nm of $SiO_2$ is deposited via Inductively Coupled Plasma Chemical Vapor Deposition (ICP-CVD) to swiftly protect the sidewall. Subsequently, holes are created using a reactive ion etching (RIE) process, facilitating the lift-off of Cr (30nm)/Au (200nm) electrodes in the ensuing step to serially connect the junctions. The electrodes exposed in the previous step are then shielded by a passivation bilayer of $SiO_2$ (200nm)/$Si_3N_4$ (200nm), with only the terminal pads being etched out by RIE for wire bonding with the PCB. Ultimately, the magnetrodes, measuring 300μm in width and featuring a tip angle of 80°, are shaped by deep reactive ion etching (DRIE). The complete process flow for the fabrication of magnetrodes is detailed in Figure 5c.

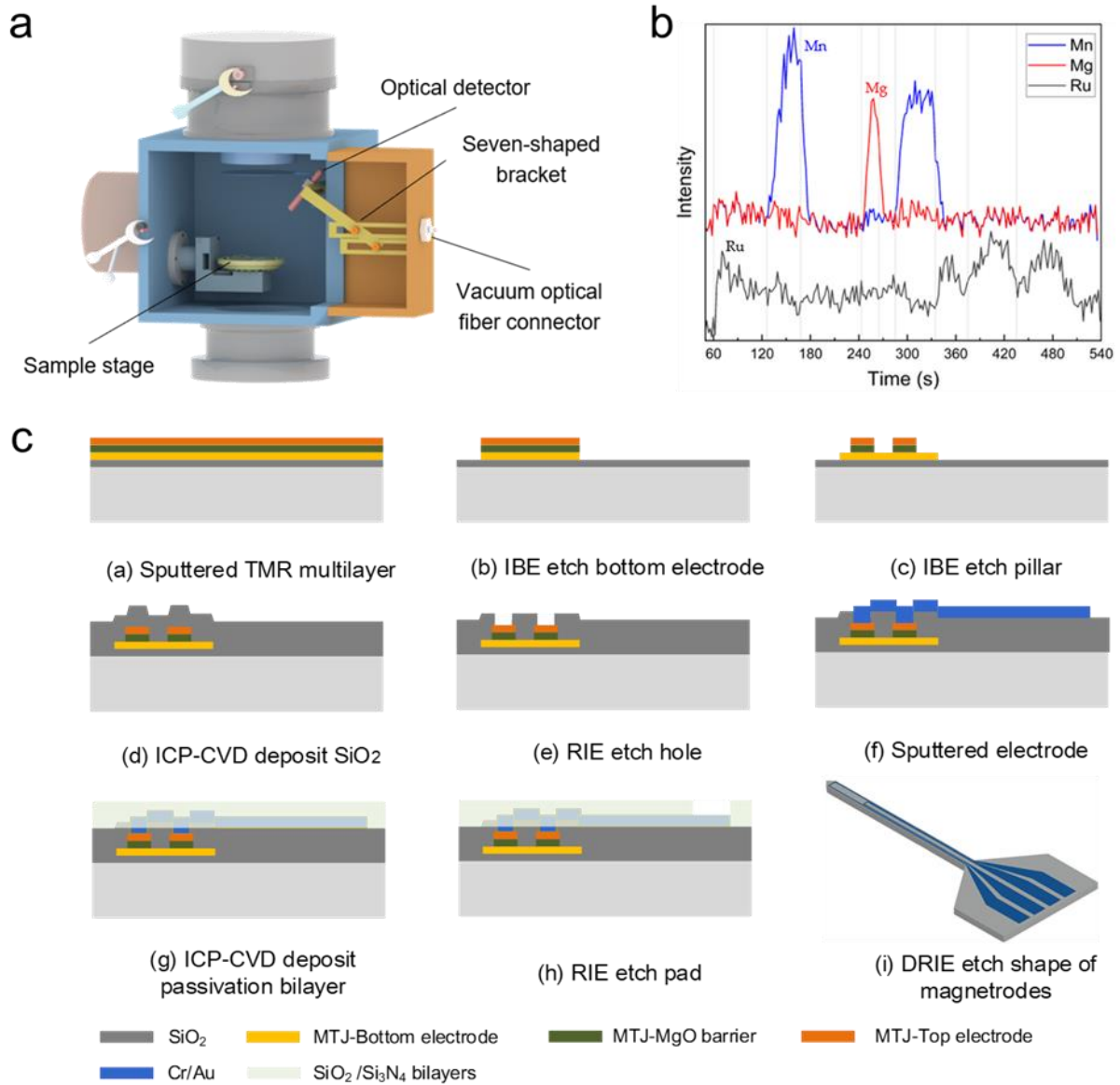

**Figure 5.** (a) The three-dimensional model of ion beam etcher; (b) Spectra monitored result during the first etching step. The rising in the spectral line indicates the commencement of etching, while the falling signifies the conclusion of the etching process. A vertical shift has been applied to the curves for better visual representation; (c)The profile of the whole process flow for magnetrodes fabrication. The $SiO_2/Si_3N_4$ bilayers are hidden in the last step.

**Set up of Measurement Systems.** Comprehensive characterization of the magnetrode necessitated two types of transport measurements: magnetotransport measurements for assessing sensitivity and MR ratio, and noise measurements to determine the LOD within the relevant frequency range. Magnetotransport measurements were conducted using the magnetic field probe station system supplied by Truth Instruments Company (Hangzhou, China). This customized system generates an in-plane magnetic field through an electromagnet, achieving ±1% uniformity at Φ2mm. The sample carrier, capable of movement in the X and Y planes, features a T-axis precision rotary stage with 360° rotation capability. During testing, the current source (Keithley 6221) powers the device, while the nanovoltmeter (Keithley 2182A) records the device voltage. These instruments are seamlessly integrated into the probe station system, complete with software for TMR testing, data storage, and export capabilities.

Accurate low-frequency noise measurement is essential for characterizing the limit of detection (LOD) in magnetic sensors. In typical experimental settings, factors like geomagnetic fields, passing vehicles, subways, and large electronic devices can induce significant fluctuations in the ambient magnetic field, thereby masking the sensor's intrinsic noise. Consequently, it is imperative to establish a specialized noise measurement system for MR sensors with a pT-level LOD. Initially, it is crucial to minimize nearby equipment and ferromagnetic objects to reduce extraneous magnetic field interference. Subsequently, a six-layer Permalloy alloy magnetic shielding barrel is customized to maximize external magnetic field interference shielding. Test results indicate that remanent magnetism within the shielding barrel's center is less than 1nT, a mere fraction (one ten-thousandth) of the external environmental magnetic field. Additionally, the system incorporates a low-noise pre-voltage amplifier (Stanford Research SR560) for filtering and amplifying the output signal, and a 16-bit acquisition card (NI 6221 or CONTEC AI-1608VIN-USB) for relaying the acquired signal to a laptop for noise power spectral density (PSD) curve analysis. When using the SR560, it is recommended to prioritize differential inputs and minimize the input cable length. This approach enhances signal integrity, reduces common-mode interference, and eliminates noise along the input signal path. To suppress industrial frequency noise, the magnetrodes, SR560, and laptop are all battery-powered. All devices within the measurement system are connected to a common ground, and all cables utilize Bayonet Neill-Concelman (BNC) shielded connectors. Figure 6 displays both the schematic and physical diagram of the noise measurement system.

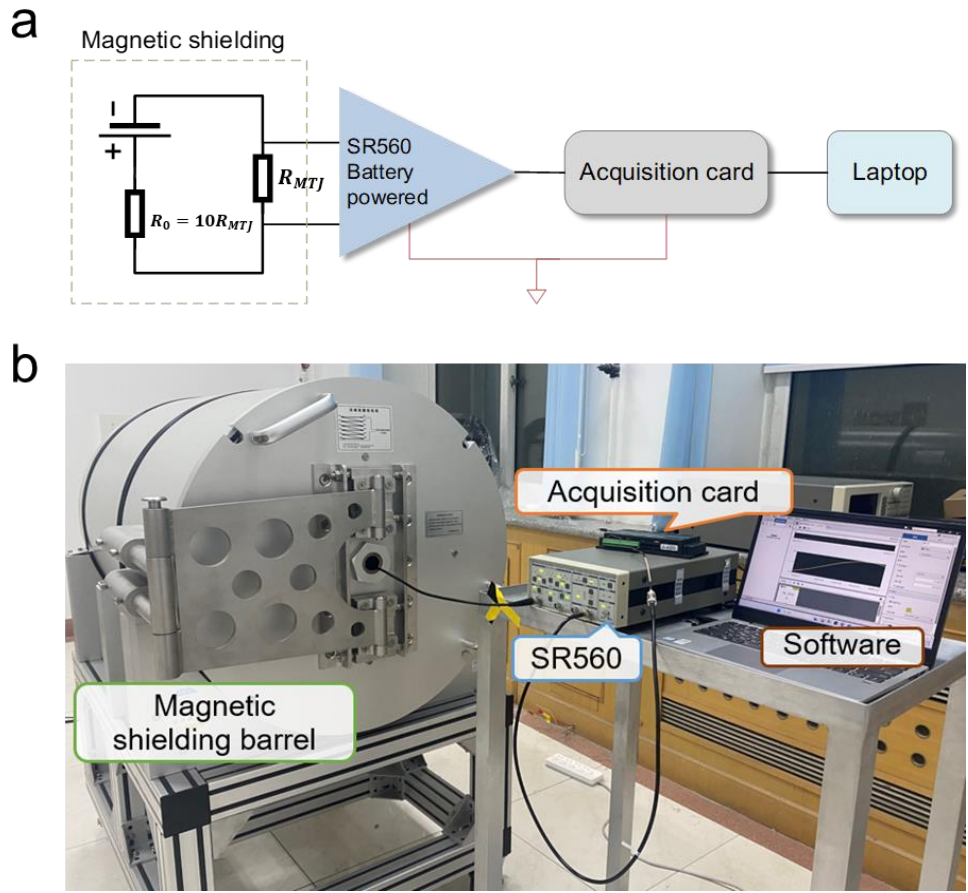

**Figure 6.** (a)Schematic diagram of the noise measurement system; (b) Physical depiction of the noise measurement system.

**Acquisition and Processing of Magnetic Signals in in vitro Simulation Experiments.** A copper wire with a diameter of 30μm simulated a neuron axon, receiving voltage signals from the neural signal simulator and generating a magnetic field. The neural signal simulator is sourced from Blackrock Microsystems, Inc., and we utilized one channel out of the available 16 channels to generate the simulated neuroelectric signal. In the experiment, the position of the copper wire is fixed, and we move the magnetrode to bring its tip sensor as close as possible to the wire. Both are placed inside a magnetic shielding barrel. Figure 7 illustrates the set up for in vitro simulation experiments. The voltage of the action potential output by the signal simulator is several tens of μVs, and the resistance of the copper wire is around 3Ω. Therefore, the current flowing through the wire is approximately a few tens of μAs. According to Biot and Savart law, assuming a distance of 20μm between the magnetrode and the wire, this would generate a magnetic field of approximately 10μT. The output signal of the magnetrode underwent filtering amplification with an SR560 filter amplifier set to a 0.3-30kHz bandpass filter and amplified 50,000 times. The signal was then real-time acquired using an NI acquisition card (NI 6221) with a sampling rate of 30kHz.

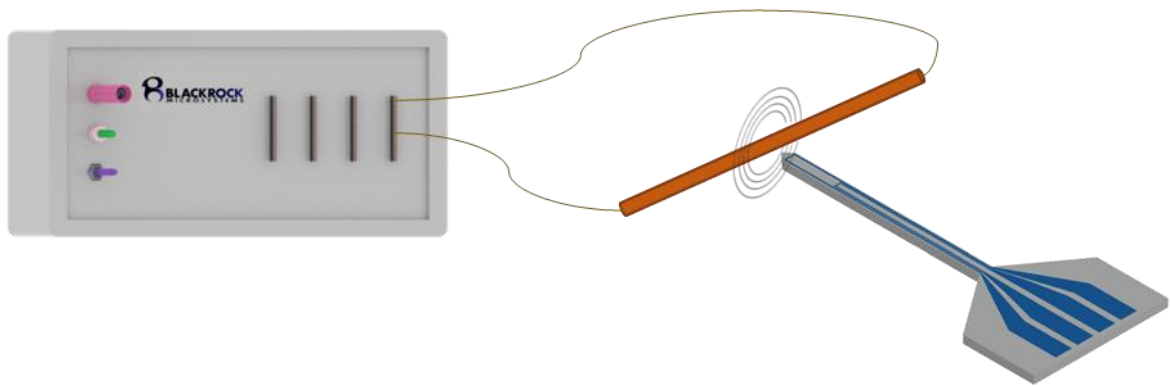

**Figure 7.** In vitro simulation experiments set up.

All data processing was then performed using MATLAB. The first step involved identifying the effective spike time points from the recorded electrical signals. The diff function was used to compute the differences between adjacent time points, resulting in a new vector "dt" representing the time intervals between each spike and the previous spike. During the iteration process, if dt was less than or equal to 1 second (the time difference threshold), no action was taken, and the next iteration was initiated. If dt was greater than 1 second, the algorithm first checked whether "i-start_idx" (current iteration position minus the start index position, i.e., the number of consecutive time points) reached or exceeded 24 (the consecutive time points threshold).

If it did, the corresponding effective spike segment was recorded in "valid_spike"; otherwise, no action was taken. The start index "start_idx" was then reset to the current iteration position, and the next iteration was initiated. The continuous time points stored in "valid_spike" were then averaged to obtain a set of spike time points.

In the second step, a notch filter was applied to remove the power line interference from the magnetic signal. The third step involved using a 250Hz high-pass filter to extract high frequency signal. In the fourth step, time windows of 1 second before and after the corresponding effective spike time points in the magnetic signal were extracted, resulting in a two-dimensional matrix. Finally, all spike events were iterated through, and FFT was performed on each spike event's data. The resulting spectra were averaged to produce spectral plots.

**ACKNOWLEDGMENTS**

This work was supported in part by the National Key R&D Program of China (2021YFB2011600), National Natural Science Foundation of China (Grant No. 62271469, 62121003), Science and Disruptive Technology Program, AIRCAS, Young Elite Scientists Sponsorship Program by CAST (No. YESS20210341), the One Hundred Person Project of the Chinese Academy of Sciences, and the Xiaomi Young Talents Program.

**References**

1. Azevedo, F.A.C., et al., Equal Numbers of Neuronal and Nonneuronal Cells Make the Human Brain an Isometrically Scaled-Up Primate Brain. Journal of Comparative Neurology, 2009. **513**(5): p. 532-541.

2. Hari, R. and R. Salmelin, Magnetoencephalography: From SQUIDs to neuroscience: Neuroimage 20th Anniversary Special Edition. NeuroImage, 2012. **61**(2): p. 386-396.

3. Barry, J.F., et al., Optical magnetic detection of single-neuron action potentials using quantum defects in diamond. Proceedings of the National Academy of Sciences, 2016. **113**(49): p. 14133-14138.

4. Caruso, L., et al., In Vivo Magnetic Recording of Neuronal Activity. Neuron, 2017. **95**(6): p. 1283-1291.e4.

5. Amaral, J., et al., Measuring brain activity with magnetoresistive sensors integrated in micromachined probe needles. Applied Physics A, 2013. **111**(2): p. 407-412.

6. Valadeiro, J., et al. Microneedles with integrated magnetoresistive sensors: A precision tool in biomedical instrumentation. in 2017 IEEE Sensors Applications Symposium (SAS). 2017.

7. Chopin, C., et al., Magnetoresistive Sensor in Two-Dimension on a 25 μm Thick Silicon Substrate for In Vivo Neuronal Measurements. ACS Sensors, 2020. **5**(11): p. 3493-3500.

8. Klein, F.J., et al., In vivo magnetic recording of single-neuron action potentials. bioRxiv : the preprint server for biology, 2023.06.30.547194. https://doi.org/10.1101/2023.06.30.547194

9. Luo, J., N. Xue, and J. Chen, A Review: Research Progress of Neural Probes for Brain Research and Brain-Computer Interface. Biosensors, 2022. **12**(12): p. 1167.

10. Kurashima, K., et al., Development of Magnetocardiograph without Magnetically Shielded Room Using High-Detectivity TMR Sensors. Sensors, 2023. **23**(2): p. 646.

11. Fujiwara, K., et al., Magnetocardiography and magnetoencephalography measurements at room temperature using tunnel magneto-resistance sensors. Applied Physics Express, 2018. **11**(2): p. 4.

12. Kanno, A., et al., Scalp attached tangential magnetoencephalography using tunnel magneto-resistive sensors. Scientific Reports, 2022. 12(1): p. 6106.

13. Amaral, J., et al., Integration of TMR Sensors in Silicon Microneedles for Magnetic Measurements of Neurons. IEEE Transactions on Magnetics, 2013. **49**(7): p. 3512-3515.

14. Yuasa, S. and D.D. Djayaprawira, Giant tunnel magnetoresistance in magnetic tunnel junctions with a crystalline MgO(0 0 1) barrier. Journal of Physics D: Applied Physics, 2007. **40**(21): p. R337.

15. Han, X., et al., High-Sensitivity Tunnel Magnetoresistance Sensors Based on Double Indirect and Direct Exchange Coupling Effect*. Chinese Physics Letters, 2021. **38**(12): p. 128501.

16. Lei, Z.Q., et al., Review of Noise Sources in Magnetic Tunnel Junction Sensors. IEEE Transactions on Magnetics, 2011. **47**(3): p. 602-612.

17. Fujiwara, K., et al., Detection of Sub-Nano-Tesla Magnetic Field by Integrated Magnetic Tunnel Junctions with Bottom Synthetic Antiferro-Coupled Free Layer. Japanese Journal of Applied Physics, 2013. **52**(4S): p. 04CM07.

18. Jin, Z., et al., Detection of Small Magnetic Fields Using Serial Magnetic Tunnel Junctions with Various Geometrical Characteristics. Sensors, 2020. **20**(19): p. 5704.

19. Gadbois, J., et al., The effect of end and edge shape on the performance of pseudo-spin valve memories. IEEE Transactions on Magnetics, 1998. **34**(4): p. 1066-1068.

20. Youfeng, Z. and Z. Jian-Gang, Micromagnetics of spin valve memory cells. IEEE Transactions on Magnetics, 1996. **32**(5): p. 4237-4239.

21. Zhu, J.-G. and Y. Zheng, The Micromagnetics of Magnetoresistive Random Access Memory, in Spin Dynamics in Confined Magnetic Structures I, B. Hillebrands and K. Ounadjela, Editors. 2002, Springer Berlin Heidelberg: Berlin, Heidelberg. p. 289-325.

22. Silva, A.V., et al., Linearization strategies for high sensitivity magnetoresistive sensors. Eur. Phys. J. Appl. Phys., 2015. **72**(1): p. 10601.

23. Friedlaender, F.J., et al., Erich Peter Wohlfarth. IEEE Transactions on Magnetics, 1991. **27**(4): p. 3469-3474.

24. Ingvarsson, S., et al., Low-Frequency Magnetic Noise in Micron-Scale Magnetic Tunnel Junctions. Physical Review Letters, 2000. **85**(15): p. 3289-3292.

25. Reed, D.S., C. Nordman, and J.M. Daughton, Low frequency noise in magnetic tunnel junctions. IEEE Transactions on Magnetics, 2001. **37**(4): p. 2028-2030.

26. Guerrero, R., et al., Low frequency noise in arrays of magnetic tunnel junctions connected in series and parallel. Journal of Applied Physics, 2009. **105**(11).

27. Schaffer, E.S., et al., A Complex-Valued Firing-Rate Model That Approximates the Dynamics of Spiking Networks. PLOS Computational Biology, 2013. 9(10): p. e1003301.

28. Zhang, J., et al., An Endpoint Detection System for Ion Beam Etching Using Optical Emission Spectroscopy. Micromachines, 2022. **13**(2): p. 259.